\documentclass[letterpaper, reqno]{amsart}
\usepackage{graphicx}
\usepackage{oldlfont}
\usepackage{citesort}

\newtheorem*{ack}{Acknowledgment}

\newcommand{\vev}[1]{\langle #1 \rangle}
\newcommand{\ket}[1]{\mid #1 \rangle}
\newcommand{\Ntil}{ N \! \! \! \! \! {}^{}_{{}_{\sim}}}

\begin{document}

\title{Gravitational Statistical Mechanics: A model}

\author{Seth A. Major and Kevin L. Setter}

\date{February 2001}
\address{Major: Department of Physics\\
Hamilton College\\
Clinton NY 13323 USA\\
Setter: Department of Physics and Astronomy\\
Swarthmore College\\
Swarthmore PA 19081 USA}

\email{Major: smajor@hamilton.edu \\ Setter: ksetter1@swarthmore.edu}

\begin{abstract}
	Using the quantum Hamiltonian for a gravitational system with
	boundary, we find the partition function and derive the
	resulting thermodynamics.  The Hamiltonian is the boundary
	term required by functional differentiability of the action
	for Lorentzian general relativity.  In this model, states of
	quantum geometry are represented by spin networks.  We show
	that the statistical mechanics of the model reduces to that of
	a simple non-interacting gas of particles with spin.  Using
	both canonical and grand canonical descriptions, we
	investigate two temperature regimes determined by the
	fundamental constant in the theory, $m$.  In the high
	temperature limit ($kT \gg m$), the model is thermodynamically
	stable.  For low temperatures ($kT \ll m$) and for macroscopic
	areas of the bounding surface, the entropy is proportional to
	area (with logarithmic correction), providing a simple
	derivation of the Bekenstein-Hawking result.  By comparing our
	results to known semiclassical relations we are able to fix
	the fundamental scale $m$.  Also in the low temperature,
	macroscopic limit, the quantum geometry on the boundary forms
	a `condensate' in the lowest energy level ($j=1/2$).
\end{abstract}

\maketitle

\section{Introduction}

The statistical mechanics of the gravitational field will provide
answers to some of the most intriguing questions on the relation
between quantum and gravitational physics.  These questions center on
key issues such as the dominance of the semiclassical states, the
states of geometry on the smallest possible scales (including
microstates of black holes), the final stages of stellar collapse, and
the early dynamics of the universe.  While general gravitational
statistical mechanics remains relatively unexplored, considerable
progress has been made in understanding quantum black holes.

In the context of string theory, many properties of extremal and near
extremal black holes have been derived.  (See Ref.  \cite{apeet} for a
review.)  Within the context of four dimensional quantum gravity there
has also been some progress.  These have included studies of canonical
ensembles for vacuum spacetimes in which one fixes the average mass of
the system and computes thermodynamic quantities using the partition
function.  For Schwarzschild black holes, such canonical ensembles were
considered in, for instance, Kastrup \cite{kastrup} and M\"{a}kel\"{a}
and Repo (who also consider the statistical mechanics of the exterior
region) \cite{mr}.  The partition function in both these cases is
divergent.  In statistical mechanics, one can introduce a chemical
potential to tame the divergence of the canonical ensemble.  This was
done by Gour for black holes with area spectra given by
integer multiples of a fundamental area \cite{gour}.

The present work makes use of canonical quantum gravity using real
connection variables \cite{barbero}, or, succinctly, ``spin net
gravity.''\footnote{By spin net gravity we mean that the discrete
structure of spin networks not only describes the states of 3-geometry
but also supports dynamics in Lorentzian four dimensional spacetime.}
Spin networks, discovered in this context by Rovelli and Smolin
\cite{RSareavol}, are linear combinations of Wilson loops based
on Ashtekar's new variables \cite{abhay}.  Spin networks are
one of the central elements of this approach to the quantization of
gravity.  There is a primer on spin networks in Ref. \cite{snp} and a review
of the general approach in Ref.  \cite{CRrev}.  The mathematics has
evolved considerably and now provides rigorous underpinnings for the
quantization \cite{AIrep} - \cite{BS} of the kinematic states of
quantum gravity.

Perhaps the most striking result of this field is the new picture it
provides of geometry on the fundamental scale.  In this picture,
3-geometry is represented by a graph.  A state of quantum geometry is
built from holonomies of the gravitational connection along a
collection of edges.  These edges are ``field lines of area'' quite
similar to electric field lines in electrostatics.  This new ``quantum
geometry'' is vastly different from our garden-variety geometry of
meter sticks, particle accelerators, and regions neighboring black
holes.  Length \cite{Tlength}, area (\cite{RSareavol,QGI,FLR}), volume
(\cite{RSareavol,lollvol,QGII,lewandowskivol,RC}), and angle
\cite{SAMangle} have been found to have discrete spectra.  In this
approach to quantum gravity, geometry is fundamentally discrete.

Within the framework of spin net gravity, there have been a number of
entropy calculations (\cite{Rbh} -\cite{abk}) which are largely based
on the ``area ensemble''.  The idea is that one uses the area, instead
of energy, as the fixed macroscopic parameter.  The resulting ensemble
is defined with respect to the area and its conjugate intensive
quantity, the ``area temperature'' or surface pressure.

We take a different approach.  Gravitational systems defined in
bounded spatial regions generically have Hamiltonians associated with 
their boundaries.  We formulate gravitational statistical mechanics using
the quantum Hamiltonian and the usual partition function: a sum over
all states of the Boltzmann weighting factor $e^{-\beta \hat{H}}$.
This is only possible when the spatial manifold has boundary.  In this
case, the Hamiltonian is the quasilocal energy associated with the
boundary.  While our statistical mechanical treatment is familiar, the
statistical sum is in stark contrast to familiar textbook
presentations of statistical mechanics, where quantized matter fields
are compared to classical geometry via the infinite volume limit.  The
ensembles we describe contain no matter nor is there a background
geometry to compare to; this is the statistical mechanics of
fundamental geometry.

Even without background geometry, we nevertheless can investigate
three, separate limits to the theory: high and low temperatures and
the macroscopic limit.  The temperature regimes are determined by the
scale of the theory, which is roughly the Planck scale.  To investigate
spatial boundaries with macroscopic areas, it is necessary to study
the model when there are a very large number of intersections between
the boundary and the underlying spin network graph.  Combining both
this macroscopic limit and the low temperature regime, we find that
the entropy is proportional to area (with logarithmic correction) and
that the system condenses in the lowest energy level.

These results come from a confluence of a number of features of this
model.  First, we make use of a quantum Hamiltonian of the system
\cite{qqe}.  It has a very simple action on spin network states.
Second, spin net gravity effectively reduces the study of quantum
gravitational field theory to quantum gravitational mechanics.  Our
model is equivalent to a system of geometric particles.  Third, there
are clearly defined scales in which the theory must match previous
results.  This enables us to fix the fundamental scale of the theory.

In the next section, we define the model that forms the basis for our
study.  We give some of the necessary classical theory before
reviewing the quantum theory and spin network states.  At each stage
we give the context to the assumptions behind the model.  These
assumptions are also collected in the final subsection
(\ref{simpassump}).  In Section \ref{can}, we study the canonical
ensemble in the hot flat space limit and define the grand canonical
ensemble in Section \ref{grand}.  This allows the study of the model
in low temperature and macroscopic limits.  Finally, the work is
summarized and some implications are discussed in Section \ref{end}.

\section{Defining the Model}
\label{modeldef}

We begin the study of general gravitational statistical mechanics with
a simple model.  To motivate the assumptions and to provide the
necessary background, we discuss the classical theory, including
boundary conditions, the quantum environment, the quantization of the
Hamiltonian, and the scale of the theory.  A summary of the model is
given in \ref{simpassump}.

\subsection{Classical Hamiltonian Theory}

We consider 4-dimensional spacetimes of the form $M = \Sigma \times
{\mathbb R}$, where the spatial region $\Sigma$ has boundary.  While
there may be more than one component of the boundary in general, such
as inner and asymptotic components, we focus entirely on a single,
2-sphere interior boundary, $\partial \Sigma$.  We do not specify
whether this is an intersection of the space with a horizon.

In the Hamiltonian formalism the classical fields on the phase space
of general relativity can be expressed using connection variables
\cite{abhay,barbero}.  These connections are the configuration
variables.  The real, $su(2)$-valued connection is $A_{a}^{i}(x)$
where $a$ is a spatial index and $i$ is the index associated to the
internal $su(2)$ space.  The conjugate momenta, or ``electric field''
$E^{ai}(x)$, is a geometric variable related to the inverse spatial
metric $q^{ab}$ by $E^{ai} E^{bi} = |q_{ab}| q^{ab}$.  The determinant
of the spatial metric is denoted $|q_{ab}|$.  The kinematic phase
space of quantum gravity is similar to SU(2) Yang-Mills theory.  The
key difference is the absence of background geometry.

Set in a spatially bounded region, classical Hamiltonian gravity
provides a clear selection criteria for the quasilocal energy: The
($3+1$)-action must be functionally differentiable.  This ensures that
histories remain within the phase space.  In general, the method of
functional differentiability generates boundary conditions on the
phase space variables, gauge parameters, and surface terms
\cite{HM,dis}.\footnote{There is some confusion in the literature on
this point.  While functional differentiability {\em can} be satisfied
when the lapse ${\cal N}$ tends to zero on the boundary, this is {\em
not} a necessary condition.  Vanishing lapse on the boundary is not
the only way in which to ensure functional differentiability of the
action.} Depending on the nature of the boundary conditions, surface
terms may be added to the action to make the theory well-defined.
These surface terms, without which the theory would be inconsistent,
are the fundamental observables.

The Hamiltonian of the theory is the surface observable arising from
Hamiltonian constraint and is enforced by the lapse function ${\cal
N}$.  This surface observable is the quasilocal energy and reduces to
the ADM energy in asymptopia \cite{HM}.  It is also the surface term
which satisfies the correct algebra: The surface term generated by the
Hamiltonian constraint satisfies the same algebra as the constraint
itself \cite{HM,dis}.  For these reasons, we use the surface term
arising from the Hamiltonian constraint as the definition of the
quasilocal energy and call it the Hamiltonian.

In the real connection variables, the variation of the connection in
the Hamiltonian constraint generates a surface term which must either
vanish when boundary conditions are applied or be canceled by the
surface term
\begin{equation}
	\label{ham}
H_{\partial \Sigma}({\cal N}) = \frac{1}{4\pi G} \int_{\partial
\Sigma} d^{2}x \epsilon^{ijk} {\cal N} n_{a} A_{b}^{i} E^{bj}E^{ak}.
\end{equation}
The lapse ${\cal N}$ has a density weight of $-1$.  When the lapse is
non-vanishing on $\partial \Sigma$, the surface term of Eq.
(\ref{ham}) must be added to the Hamiltonian constraint.  This
classical expression reduces to the quasilocal energy of Brown and
York \cite{BY}, to the ADM energy in asymptopia, and to the
Misner-Sharp mass in spherical symmetry \cite{HM}.
This Hamiltonian is the surface term which ensures functional
differentiability of Lorentzian general relativity in terms of real
connection variables.

With the addition of the Hamiltonian of Eq.  (\ref{ham}) the boundary
conditions required by functional differentiability of the Hamiltonian
constraint are simply expressed.  Let $\kappa^{abi} = {\cal N}
\epsilon^{ijk} E^{aj} E^{bk}$, then $\delta n_{a}
\kappa^{abi}|_{\partial \Sigma} = 0$.  However, a complete set of
boundary conditions includes kinematic boundary conditions, which
arise from functional differentiability of the Gauss, diffeomorphism,
and Hamiltonian constraints, and dynamic boundary conditions.  Dynamic
boundary conditions ensure that the boundary conditions are preserved
under evolution.  These are found by computing the constraint algebra.
See Ref.  \cite{HM} (or, for a more general analysis, Ref.
\cite{dis}) for details.

One set of kinematic boundary conditions which ensures functional
differentiability of the constraints with the surface term of Eq.
(\ref{ham}) is
\begin{equation}
	\begin{split}
		\label{kbcs}
		&\delta (n_{a} \epsilon^{ijk} E^{aj}
		E^{bk})|_{\partial \Sigma} = 0 \\
		&\delta {\cal N}|_{\partial \Sigma} =0 \text{, the
		lapse is fixed, and} \\
		&N^{a}|_{\partial \Sigma}=0 \text{, so that spatial
		diffeomorphisms vanish on the boundary.}
	\end{split}
\end{equation}
To ensure that the theory is dynamically well-defined one must impose
further boundary conditions on $\partial \Sigma$.  One consistent
set is
\begin{equation}
	\begin{split}
		\label{dbcs}
		&D_{b} \kappa^{abi}|_{\partial \Sigma} = 0 \\
		&\Lambda^{i}|_{\partial \Sigma} = 0 \\
		&\partial_{a} {\cal N}|_{\partial \Sigma} = 0 \text{
		so that the lapse is fixed to be a constant on }
		\partial \Sigma.
	\end{split}
\end{equation}
We choose the constant lapse to be 1.

Both sets of boundary conditions, Eqs.  (\ref{kbcs}) and
(\ref{dbcs}), comprise a complete, consistent set of boundary
conditions for the model.  As is clear from these conditions, the
model is defined for hypersurface orthogonal spatial boundaries with
fixed electric field. Finally, we note that these boundary
conditions are not in the category of boundary conditions which
includes isolated horizons.  Isolated horizons require that the lapse
vanish on the horizon \cite{abk}.

\subsection{Quantum Domain}

To investigate quantum statistical mechanics, one must investigate
both the Hilbert space of the bounded theory and the action of the
quantum Hamiltonian.  We give an outline of the state space framework
here though the reader is encouraged to refer to references
\cite{AIrep,ALrep,baez,baez2,ALproject1,ALProj,ALMMT,MM,BS} for the
elegant theory behind this sketch.  The quantum degrees of freedom are
captured by holonomies along paths embedded in $\Sigma$.
Gravitational analogs of Wilson loops, a holonomy of the connection
gives an element of SU(2).  This simple observation is the key to the
quantum configuration space.  One defines a generalized connection to
be a map from each path to the group.  Like the holonomy of the
classical connection, the generalized connection describes parallel
transport along paths.  It is convenient to describe a collection of
paths by a set of edges of a single graph.  With the definition of
generalized connections and a graph, one can introduce a measure on
the space of generalized connections simply by using the $SU(2)$ Haar
measure on every edge.  It turns out that the space of generalized
connections is the completion of the space of smooth connections
\cite{AIrep,ALrep}.  This space is the quantum configuration space.  A
basis on this space is described by spin networks \cite{baez}.

An open spin network $s$ consists of the triple $( {\sf G}, {\bf n,
i})$.  The graph ${\sf G}$ is embedded in the 3-dimensional space and
is comprised of a set of edges and vertices.  Each edge is assigned
a nontrivial, irreducible representation of $SU(2)$.  On
every edge $e$ this representation is labeled by an integer $n_{e}$
from the set ${\bf n}$ where $n=2j$.  The minimum representation, or
edge label, is $n=1$.  The edges intersect at vertices.  The number of
edges incident to a vertex is called the valence.  In this model
valence-one vertices, the ``open edges,'' only occur at the
intersection of the graph with the boundary.  Intertwiners, ${\bf i}$,
assign to each vertex $v$ a vector $i_{v}$ in the tensor product of
the representations labeling the edges incident to the vertex $v$.  If
a vertex is in the interior, the vector $i_{v}$ is invariant under the
action of $SU(2)$.  If the vertex lies on the boundary, $i_{v}$ is not
invariant under $SU(2)$.  We label these boundary intertwiner vectors
by an integer which corresponds to the $m_{j}$ values familiar from
angular momentum.

In more picturesque language, edge labels record the number of
spin-$1/2$ ``threads'' in a single edge of a spin network. 
Intertwiners label the ways in which all incident threads are
connected.  Vertices at the ends of the open edges are labeled by the
state of the spin-$1/2$ threads.

Notice that any open spin network determines a set of valence-one
vertices, which determine a Hilbert space given by a product of
angular momentum states $\ket{j \, m_{j}}$.  These are the states on
the boundary of the space that determine the quasilocal energy.

We base the statistical mechanics on the quantization of the boundary
Hamiltonian of Eq. (\ref{ham}).  Using loop techniques and recoupling
theory, the boundary Hamiltonian of Eq.  (\ref{ham}) can be quantized
\cite{qqe}.  The quantization of the Hamiltonian is not free of
choices.  However, the most general form of the operator is, for
transversal edges,
\begin{equation}
	\label{qham}
	\hat{H}_{\partial \Sigma} \ket{s} = m \sum_{v \in \partial
	\Sigma \cap {\sf G}} {\cal N}_{v} \sqrt[4]{n_{v}(n_{v}+2)} \ket{s}.
\end{equation}
This operator depends only on vertices in the boundary.  The lapse ${\cal
N}_{v}$ is the scalar (vanishing density weight) lapse at the vertex
$v$.  In light of the boundary conditions of Eq.  (\ref{dbcs}),
the lapse is constant on the boundary.  We set ${\cal N}_{v}=1$ for
all boundary vertices $v$.

There are two properties of the Hamiltonian operator of Eq.
(\ref{qham}) which impact the statistical mechanics.  First, each edge
contributes to the energy independently of the other edges.  This
simplifies the model considerably, as the total quasilocal energy is
the sum of the contributions of each edge.  Thus, as we will see,
the problem reduces to a gas of nearly identical particles.
Second, as is clear from the form of the Hamiltonian operator of Eq.
(\ref{qham}), the energy of the system does not depend on the vectors
of the boundary vertices; the $m_{j}$ value does not count.  Thus, the
system is degenerate.

There is an important caveat to this degeneracy.  We define the model
only through the component of the system which determines the energy,
the boundary theory.  If one were to include considerations of the
interior state, one would include the vast degeneracy, presumably
infinite, of the interior intertwiner bases, knot states of the edges,
and other diffeomorphism invariant aspects of the spin network graph
\cite{GR}.

In statistical mechanics, the macroscopic behavior depends critically
on the particle statistics.  In our model the intersections of the
surface with the network define a set of independent, distinguishable
particles with spin.  We take particles with different spin to be
distinguishable.  To motivate this it is best to describe measurements
on the fundamental geometry.

In quantum geometry, each intersecting edge determines the `local'
shape of the surface.  Not only does a label $n$ determine the area
contribution of a single edge, but the vertex also can determine the
local curvature, as in Ref.  \cite{abk}.  As observed from the
interior, the labels on these edges - the amount of geometric flux -
are measurable via the area operator.  Fluctuations in these labels
are, in principle, observed as localized sources of radiation.  The
expectation is that, even if the macroscopic geometry is symmetric,
the quantum geometry will be fluctuating on the microscopic scale.
The localization can be achieved with the spin network itself if the
spin network is sufficiently complex.  Thus, simply based on the
abstract graph, its labels and its intertwiners, we expect that it is
possible to distinguish spin network edges and their fluctuations.
This result may also be derived by examining the role of
diffeomorphisms on the boundary \cite{Rbh}.

With these considerations the following picture emerges:
Gravitational statistical mechanics of the model is based on a set
of non-interacting particles.  Two geometric particles are
distinguishable if they have different spins.  The particles induced
from an edge with label $n$ have a degeneracy of $n+1$.

\subsection{Scale, Parameters, and Summary}
\label{simpassump}

The constant $m$ in Eq.  (\ref{qham}) sets the fundamental scale of
the theory.  It is the same scale as that of the length $\ell$ which
appears in the geometric observables.  For instance, the area
contribution of a single edge $e$ is given by $\ell^{2}
\sqrt{n_{e}(n_{e}+2)}$.  The two parameters $m$ and $\ell$ are
inversely related with $m = \hbar / \ell $.

We retain the overall scale parameter in terms of $m$ or $\ell$. 
There are two reasons for doing this.  First, we wish to derive the
relation between the fundamental scale of the model and the Planck
length $l_{P} := \sqrt{\hbar G/c^{3}}$.  Second, this enables us to
address a subtlety arising in the definition of the connection.  As
Immirzi has emphasized, in the canonical transformation used to define
the connection variables there is a family of choices generated by one
non-zero, real parameter $\gamma$, ${}^{\gamma}A_{a}^{i} =
\Gamma_{a}^{i} - \gamma K_{a}^{i}, {}^{\gamma}E^{ai} = (1/
\gamma)E^{ai}$ \cite{I}.  We absorb $\gamma$ into the constants $m$
and $\ell$.  Finally, we retain Boltzmann's constant $k$ throughout.

Since $m$ sets the scale, the regimes of interest are determined
relative to this constant.  The significant dimensionless parameter is
the ratio of this energy scale to the thermal energy, $\beta m$.
Temperatures of order unity ($T \sim 1/km$) are roughly $10^{32}$ K,
so garden-variety energy scales of the universe are far below the
natural scale of the theory; for quantum geometry, the universe is a
chilly place.  Thus the more realistic regime is the low temperature
limit.  Throughout this paper we also use such terms as
``microscopic'' and ``macroscopic.''  By microscopic we mean
boundaries with area on the order of $\ell^{2}$.  By macroscopic we
mean boundaries with areas much greater than $\ell^{2}$, i.e.
boundaries on scales of elementary particles up through astrophysical
objects.  As we see in Section \ref{grand}, to use microscopic results
in the macroscopic regime requires a study of the scaling of physical
quantities as they diverge.  All semiclassical results need to be
closely examined in low temperature, macroscopic limits.

When we compare the thermodynamics of this model with semiclassical
results, we fix the fundamental scale of the theory.  This is done in
Section \ref{thermo}, where we compare the area-entropy relationship
of our model to the Bekenstein-Hawking area-entropy relation.

In summary, the remainder of the paper deals exclusively with the
model defined by:
\begin{enumerate}
	\item A spatial manifold having boundary $\partial \Sigma$.
	\item We assume that the quantum boundary conditions are a
	straightforward implementation of the classical boundary
	conditions of Eqs.  \ref{kbcs} and \ref{dbcs}.
	\item A quantum Hamiltonian given by Eq.  (\ref{qham}) with:
	\begin{enumerate}
	\item Fixed, unit lapse on $\partial \Sigma$,
	\item A degeneracy of $n_{e}+1$ associated to every edge $e$.
	\end{enumerate}
	\item We assume valence-one boundary vertices, i.e. the
	intersections ${\sf G} \cap \partial \Sigma$ occur along single
	edges
	\item We assume edges with different spins can be distinguished.
\end{enumerate}
Hence, the model reduces to a non-interacting, degenerate gas of
distinguishable particles with spin.

\section{Canonical ensemble}
\label{can}

This section provides groundwork for both the canonical and grand
canonical ensembles.  In the canonical ensemble, we concentrate on the
high temperature limits of the theory.

Given any boundary with non-vanishing area, a set of edges will
intersect the boundary.  In the canonical ensemble, we fix the total
number of intersections between the spin network graph and the
boundary $\partial \Sigma$ to be $N$, while allowing the energy of the
system (the labels on the edges) to fluctuate.  These intersecting
edges are the only edges in the spin network that are open.

We investigate the statistical mechanics of $N$ distinguishable
non-interacting geometric particles.  To specify a state of the
system, we need two things: a list of edge labels $\{n_i\}$ for those
edges that intersect the boundary $\partial \Sigma$, and an ``$m_{j}$"
value for each of these edges.  Since the particles are
distinguishable, $\{n_i\}$ will be an ordered list.

The canonical partition function, $Z= \text{Tr } e^{-\beta \hat{H}}$,
becomes
\begin{equation}
	\label{canpart1}
	Z_{N}(T) = \sum_{\{ n_{i}\}} g(\{n_{i}\}) \: e^{-\beta
	(\epsilon_{n_1} + \dots + \epsilon_{n_N})}
\end{equation}
with the quantum Hamiltonian of Eq.  (\ref{qham}) with unit lapse.
The energy contribution of the $i$th edge is given by $\epsilon_{n_{i}}
:= m \sqrt[4]{n_{i}(n_{i}+2)}$, where $m$ is the fundamental mass scale
for the system.  The degeneracy due to the boundary
vertices is  
\begin{equation}
	\label{degen}
	g(\{n_i\}) = \prod_{i=1}^{N} (n_{i} + 1).
\end{equation}
As the particles are independent, the partition function reduces to
\begin{equation}
	Z= \left( \sum_{n=1}^{\infty} (n+1) e^{-\beta \epsilon_{n}}
	\right)^{N}.
\end{equation}
With this partition function it is an easy matter to compute the
thermodynamic potential
\begin{equation}
	{\cal A} = - \frac{1}{\beta} \ln Z = - \frac{N}{\beta} \ln f,
\end{equation}
where
\begin{equation}
	\label{fdef}
	f:= \sum_{n=1}^{\infty} (n+1) e^{- \beta \epsilon_{n}}
\end{equation}
is the one particle partition function.

Normally at this stage one would study the behavior of the partition
function and thermodynamic quantities in the infinite volume limit. 
As we are studying the underlying geometry itself, we cannot do this. 
We have no background for comparison.  Nonetheless, we can explore the
high temperature limit.

In the high temperature limit, $\beta m \ll 1$, the sum in $f$ may be
approximated by an integral\footnote{More precisely, one can establish
upper and lower bounds on the integral using the Cauchy-Maclaurin
integrals.  In the high temperature limit, both these bounds have the
leading order behavior displayed in Eq.  (\ref{hightempf}).} which can
be integrated exactly
\begin{equation}
	\begin{split}
f & \approx  \int_{1}^{\infty} (n+1) e^{-\beta \epsilon_{n}}\\
&= \frac{2}{(\beta m)^{4}} \, e^{- \sqrt[4]{3} \beta m}
\left( 6 + 6 \sqrt[4]{3} \beta m + 3 \sqrt{3} (\beta m )^{2} +
\sqrt[4]{3^{3}} (\beta m)^{3} \right).
\end{split}
\end{equation}
The single particle partition function behaves as
\begin{equation}
	\label{hightempf}
     f \approx \frac{12}{(\beta m)^{4}} - \frac{3}{2} +  O( \beta m)
	\approx \frac{12}{(\beta m)^{4}}
\end{equation}
for high temperatures.  In this temperature regime, physical
quantities are easy to compute.  The entropy is
\begin{equation}
	\label{canent}
	S = - \frac{\partial {\cal A}}{\partial T} = k N \left( \ln 12 - 4
	\ln (\beta m) + 4 \right).
\end{equation}
The heat capacity simply scales with the number of intersections
\begin{equation}
	C = T \frac{\partial S}{\partial T} = 4 k N.
\end{equation}
As $C>0$, Le Ch\^atelier's principle is satisfied and the system is
thermally stable.  One initially would expect that since the behavior
of Hawking radiation requires that the specific heat of black holes to
be negative, this high temperature result would preclude black holes.
However, there are quite general arguments \cite{louko} that the
specific heat of Lorentzian black holes ought to be positive.
Further, York \cite{york} found that, using Hawking's result for
spherically symmetric black holes, black holes in a box can be
thermodynamically stable.  In fact, York found that there were two
possible mass solutions, with one stable and one unstable.  He
described these as two different phases of the system, one with stable
black holes and one of hot flat space.  To map out the correspondence
between the model York considers and the present model, we would need
to add an additional boundary and a parameter which measured the size
of the outer boundary.  The topology of our present model prevents us
from seeing both solutions.

Average energy and area are also easy to compute.  The
average energy is given by
\[
\vev{E} := \sum_{i=1}^{N} \vev{E_{i}}
\]
with
\[
\vev{E_{j}} = \frac{1}{f} \sum_{n_{j}=1}^{\infty}
\epsilon_{n} e^{-\beta \epsilon_{n}}.
\]
Since all the expectation values for each of the particles are all
identical, $\vev{E}=N \vev{E_{i}}$ for any $i$.  In the high
temperature regime we make the approximation
\[
\vev{E_{i}} \approx \frac{m}{f} \int_{1}^{\infty} dx (x+1)
\sqrt[4]{x(x+2)} e^{-\beta m \sqrt[4]{x(x+2)}}.
\]
The integral can be done exactly.  The leading order behavior
gives a total average energy of
\begin{equation}
	\label{vevehightemp}
	\vev{E} = 4 N k T
\end{equation}
in the high temperature limit.  Similarly, the area of the bounding
2-surface, in the high temperature limit, is given by
\begin{equation}
	\label{vevarea}
	\vev{A} = 20 \ell^{2} N \left(\frac{kT}{m} \right)^{2} .
\end{equation}
All of these physical quantities are proportional to the number of
intersections $N$.  From the entropy (\ref{canent}) and the area
(\ref{vevarea}) results, we see that the entropy and area obey the
Bekenstein-Hawking proportionality $S= k \tfrac{ A}{4 \ell^{2}}$ only
at specific temperatures.  In this high temperature approximation,
this occurs when $\beta m \approx 0.78$ and $5.8$.  Since the
Bekenstein-Hawking result is valid in lower temperatures ($\beta m \gg
1$), these are spurious solutions.

It is perhaps not surprising that semi-classical physics differs with
these results at high temperature scales.  We should not be surprised
to discover that the semi-classical calculations, derived in the low
temperature domain, do not agree with results derived for temperatures
much greater than the Planck scale!

As is often the case in statistical mechanics, to investigate lower
temperatures it proves easier to work with the grand canonical
ensemble.  This is what we turn to next.

\section{Grand Canonical Ensemble}
\label{grand}

In the statistical mechanics of particles the grand canonical ensemble
includes varying particle number.  In this case we allow the number of
intersections between the spin network state and the boundary to
fluctuate.  This means that we consider the category of dynamics in
which spin network states evolve by changing the underlying graph in a
manner consistent with kinematical constraints and boundary
conditions.\footnote{Note that the classical boundary conditions, Eqs.
(\ref{kbcs}) and (\ref{dbcs}) allow the boundary metric, including
``area density'' $n_{a} E^{ai}$, to change.}

A new element in our approach is the use of the chemical potential. 
Though used in the Schwarzschild case in Ref.  \cite{gour}, the
chemical potential has not been used within the context of spin net
gravity.  Allowing the number of particles to fluctuate provides a
number of benefits.  The scale of the theory, $m$, is so large that,
in the canonical ensemble, any change of the system is fantastically
unlikely for temperatures such as occur in our universe.  The dynamics
appears to be ``frozen''.  Indeed, the analysis would seem to require
that all geometric objects have only {\em two} spin-$1/2$ edges!  The
chemical potential allows interesting behavior to occur at lower
temperatures and for macroscopic configurations.  Further, the
chemical potential provides a measure of the geometric interaction
between the interior and the boundary.

Calling the number of intersections, $N$, and the chemical potential,
$\mu$, the resulting grand canonical partition function is
\begin{equation}
	{\cal Z}(T, \mu) = \sum_{N=1}^{\infty} \sum_{\{n_{i}\}}
	g(\{n_{i}\}) e^{-\beta \left(\epsilon_{n_1} + \dots +
	\epsilon_{n_N}- \mu N \right)},
\end{equation}
where, as before, $g(\{n_{i}\})$ is the degeneracy of the edge states.
Since we consider boundaries with non-vanishing area, the sum over $N$
begins at $1$.\footnote{Readers familiar with spin networks will see
that a single edge intersection would violate the admissibility
conditions, in particular the evenness condition.  One can account for
this in the ensemble without affecting the physical results presented
here \cite{kevinthesis}.} Each edge independently contributes to
energy, so the partition function simplifies.
\begin{eqnarray}
	{\cal Z} &=& \sum_{N=1}^{\infty} \left( \sum_{n=1}^{\infty} (n+1)
	e^{-\beta (\epsilon_{n} - \mu}) \right)^{N} \\ \nonumber
	&=& \frac{fz}{1 - fz}
\end{eqnarray}
in which the fugacity $z$ is defined as usual, $z = e^{\beta \mu}$,
and $f$ is defined in Eq.  (\ref{fdef}).

This partition function diverges as $fz$ approaches 1.  There is
physics in this divergence.  To investigate the divergence we first
compute some physical quantities.

\begin{figure}
	\begin{center}
\includegraphics[scale=0.6]{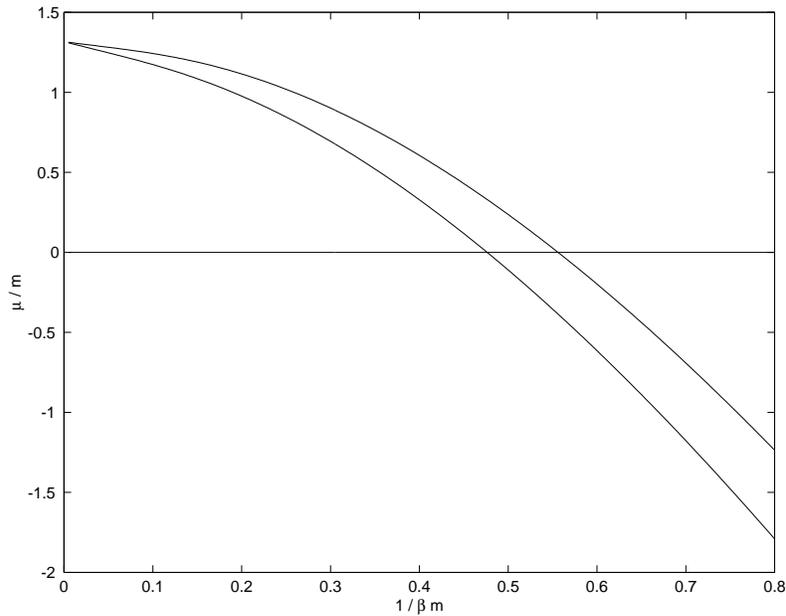}
\caption[Chemical potential as a function of temperature]{
\label{nisone}
The restriction that the average number of intersections be greater or
equal to 1, $\vev{N} \ge 1$, yields a relation between the temperature
and the chemical potential.  The upper curve is the plot of the
relation, $\vev{N} = 1$.  There is a maximum temperature for every
chemical potential.  The lower curve, or ``macroscopic curve,'' is the
relation $fz=1$ which corresponds to infinite intersection number. 
The curve near zero temperature is roughly linear, with a slope of
about $-0.7$ and an intercept of $\sim \sqrt[4]{3}$.  Both curves have
an error on the order of $10^{-3}$ at the Planck temperature $\beta
m=1$.}
\end{center}
\end{figure}

The thermodynamic potential is given by
\begin{equation}
	\Omega = - kT ( \ln fz - \ln (1-fz)).
\end{equation}
The ensemble average of the number of intersections is
\begin{equation}
	\vev{N} = - \frac{\partial \Omega}{\partial \mu}
	= \frac{1}{1-fz}.
\end{equation}
For physical configurations, the ensemble average $\vev{N}$ must be
greater than or equal to 1.  The lower bound, $\vev{N} = 1$,
introduces a relation between temperature and chemical potential,
\[
\mu = - \frac{\ln f}{\beta} ,
\]
which selects physical configurations when the boundary has
non-vanishing area.  A plot of the relation $\vev{N} = 1$ is given in
Fig.  (\ref{nisone}).

The number of intersections diverges when the product $fz$ approaches
1.  Since $f$ is a monotonically increasing function of the
temperature, these restrictions yield a maximum temperature for every
chemical potential.  This relation is the upper curve in Fig.
(\ref{nisone}).  Temperatures higher than shown in the figure send the
partition function to infinity and $\vev{N}$ to unphysical, negative
values.  All physical configurations lie inside the region between the
two curves in Fig.  (\ref{nisone}).  We call this region bounded by
the curves $\vev{N} = 1$ and $fz \simeq 1$ the ``macroscopic band.''

To better understand the geometry at the divergence, when $fz
\rightarrow 1$, it is worth deriving the area of the bounding surface.
The area may be obtained through the simple relation
between the Hamiltonian and the area at each intersection $i$
\cite{qqe}
\begin{equation}
	\frac{\epsilon_{i}}{m} = \frac{\sqrt{a_{i}}}{\ell^{2}}.
\end{equation}
A short calculation shows that
\[
\vev{A} = \frac{1}{{\cal Z}} \sum_{N=1}^{\infty} N (fz)^{N}
\frac{1}{f} \sum_{n=1}^{\infty} a_{n} (n+1) e^{- \frac{\beta m}{\ell}
\sqrt{a_{n}}}.
\]
Thus,
\begin{equation}
	\label{area}
\vev{A} = \vev{N} \vev{a},
\end{equation}
in which $\vev{a}$ is the single particle average area.

Since the area scales with $\vev{N}$, macroscopic areas occur for
large $\vev{N}$.  The divergence $\vev{N} \rightarrow \infty$ when $fz
\rightarrow 1$ is physical in that the divergence selects the region of
parameter space where macroscopic objects live.  In this model of
discrete space, macroscopic objects are found in $(1/\beta m, \mu /
m)$ space between the curves shown in Fig.  (\ref{nisone}).

At low temperatures, it is easy to characterize the macroscopic
limit.  When $\beta m \gg 1$, $fz$ may be
expressed as
\begin{equation}
	\label{fzlowtemp}
fz = 2e^{-\beta (\epsilon_{1} - \mu)} \left( 1 +
\sum_{n=2}^{\infty} \left( \frac{n+1}{2} \right) e^{- \beta
(\epsilon_{n} - \epsilon_{1})} \right) = 2e^{-\beta
(\epsilon_{1} - \mu)} + O( e^{- \beta m \Delta}),
\end{equation}
where $\Delta := (\sqrt[4]{8} - \sqrt[4]{3})$ is the numerical
difference in the first two energy levels.  Neglecting the higher
order terms, the macroscopic condition $fz \simeq 1$ gives us
\begin{equation}
	\label{mumacrolowtemp}
\mu \simeq - \frac{\ln 2}{\beta} + \epsilon_{1}.
\end{equation}
This relation, the low temperature, leading order approximation to the
top curve in Fig.  (\ref{nisone}), proves useful when comparing the
entropy and area.  We continue to use the ``$\simeq$'' whenever we
express quantities in the low temperature, macroscopic limit.

\subsection{Condensation}

By examining the occupation numbers of the edge labels, one can find
macroscopic occupation of the lowest level.  If one rewrites the
partition function in terms of number of intersections at the $k$th
level, $N_{k}$, then one quickly finds the average
\[
\vev{N_{k}} = - \frac{1}{\beta} \frac{\partial \ln {\cal Z}}{\partial
\epsilon_{k}}  = \frac{(k+1) e^{- \beta \epsilon_{k}}}{f(1 - fz)}.
\]
As might be expected, these quantities diverge with $\vev{N}$ for all
$k$.  It is more interesting to investigate how the relative
proportions $\vev{N_{k}}/{\vev{N}}$ behave.  The ratio is
\begin{equation}
\frac{\vev{N_{k}}}{\vev{N}} = \frac{k+1}{f} e^{- \beta \epsilon_{k}}
\end{equation}
which, curiously, is independent of the fugacity $z$.  With the same
techniques just employed in the limit for $fz$ in Eq.
(\ref{fzlowtemp}), at low temperatures the occupation of the lowest
possible level, $k=1$ behaves as
\begin{equation}
	\label{Ncond}
\frac{\vev{N_{1}}}{\vev{N}} \simeq 1 - O( e^{- \beta m \Delta}).
\end{equation}
As $\vev{N}$ becomes macroscopic, so does $\vev{N_{1}}$. The
system ``condenses'' in the spin-$1/2$ state, meaning that in the limit
of large $\vev{N}$ and at low temperatures, the occupation of the
lowest state accounts for virtually all the particles.  This is much
like Bose-Einstein condensation, which is characterized by a finite
fraction of states in the zero momentum state in the infinite particle
limit
\[
\lim_{N \rightarrow \infty} \frac{\vev{n_{0}}}{N} > 0.
\]
If the ratio vanishes in the limit, then there is no condensation. 
While the analogous ratio in the quantum gravity case, Eq. 
(\ref{Ncond}), is positive and near unity, it is difficult to be
precise as the $\vev{N} \rightarrow \infty$ condition is the
macroscopic limit and results in a more complicated relation between
$\mu$ and $\beta$.  (The condition $fz \simeq 1$ ties the chemical
potential to the temperature, producing the macroscopic curve.) 
Nevertheless, it is easy to characterize the system in the low
temperature limit; the model relaxes exponentially into the lowest
state as the ratio of neighboring low lying states
\[
\frac{\vev{N_{2}}}{\vev{N_{1}}} = \frac{3}{2} e^{- \beta m
(\sqrt[4]{8} - \sqrt[4]{3})}
\]
vanishes exponentially as $T \rightarrow 0$.

There is another difference between the condensation in this model and
in a Bose-Einstein system.  The fluctuations, $\vev{\Delta N^{2}} =
\vev{N^{2}} - \vev{N}^{2}$, in intersection number behave as
\[
\vev{\Delta N^{2}} = \frac{fz}{(1-fz)^{2}}.
\]
Hence, the relative dispersion $\vev{\Delta N^{2}}/\vev{N}^{2}$ goes
to 1 in the low temperature, macrcoscopic limit.  The relative
dispersion of the number of intersections at the lowest level also
behaves this way.  In a Bose-Einstein system, the analogous relative
dispersions would vanish at low temperatures.  In our model, despite
the fact that the spin-$1/2$ state is macroscopically occupied, we
still have a tempest of fluctuations.  As discussed in Section
\ref{end} these fluctuations lead to potentially observable effects.

\subsection{Gravitational thermodynamics}
\label{thermo}

To check whether this model of quantum gravitational statistical
mechanics reproduces the results of gravitational thermodynamics, we
first derive the remaining macroscopic parameters of the system,
entropy and average energy.  Since we have not specified the boundary
conditions so carefully as to pick out a black hole solution, it is
interesting to see whether the relations of black hole mechanics are
included in our thermodynamics.  The remaining physical quantities are
derived in a straightforward way, using standard statistical
mechanics.

Defining the ensemble average of energy of a single intersection as
\[
e := \sum_{n=1}^{\infty} \epsilon_{n} (n+1) e^{- \beta \epsilon_{n}},
\]
the energy is given by
\begin{equation}
	\vev{E} = - \frac{\partial \ln {\cal Z} (T, z)}{\partial \beta} =
	\vev{N} \frac{e}{f}.
\end{equation}
The entropy is given by
\begin{equation}
	\label{entropy}
	S = - \left( \frac{\partial \Omega}{\partial T} \right)_{\mu} =
	k \ln \vev{N} + k \ln (fz) + \frac{1}{T} \vev{E} - \frac{\mu}{T}
	\vev{N}.
\end{equation}

The first thermodynamic relation we study is the Bekenstein-Hawking
formula $S = k A/4 l_{P}^{2}$.  A quick glance at the equations for
area, entropy, and energy in this model show a more complicated
dependence than this simple proportionality.  However, this is to be
expected as the thermodynamic result holds at low temperatures and
macroscopic geometries.  At low temperatures, the area of Eq.
(\ref{area}) is simply $\vev{N}$ times the basic area element,
\[
\vev{A} \simeq \ell^{2} \sqrt{3} \vev{N},
\]
so that the area scales with the number of intersections.

Using the relation between chemical potential and $\beta$ at low
temperatures, Eq.  (\ref{mumacrolowtemp}), we find that
\begin{equation}
	\label{bekenstein}
	S \simeq k \frac{ \ln 2}{\sqrt{3}} \frac{\vev{A}}{\ell^{2}} + k
	\ln \left( \frac{\vev{A}}{\ell^{2}} \right)
\end{equation}
in the low temperature, macroscopic limit.  This establishes the
proportionality between entropy and area, with a logarithmic
correction.  This is identical to the result for black holes!

The reason for the proportionality factor $\ln 2 / \sqrt{3}$ is easy
to see.  Suppose the surface area is quantized with uniformly spaced
levels $A = n \alpha l^{2}$, where $n=1,2, \ldots$, $\alpha$ is a
fixed parameter, and $l$ is the fundamental length.  This suggests
that the bounding surface is like a quilt sewn together of Planck
scale patches, each of area $\alpha l^{2}$.  If every patch has two
possible states, then a surface of area $A=n \alpha l^{2}$ has $2^{n}$
surface states.  This degeneracy contributes to the entropy and $S =
\ln 2^{n} = \ln2 (A/ \alpha l^{2})$.  This is precisely what we have
in Eq.  (\ref{bekenstein}), with $\alpha = \sqrt{3}$.  Thus, the
entropy is equal to the log of the number of states of the lowest
level times the number of units of the fundamental area element
$\sqrt{3} \ell^{2}$.

In the context of non-perturbative quantum gravity, a logarithmic
correction has been derived before by Kaul and Majumdar \cite{km}.
However, they found that the correction term was negative, indicating
that the semiclassical result was the maximal entropy.  In other
derivations of the logarithmic term, the corrections arise from
quantum effects involving matter fields on a classical background.

To reproduce the thermodynamic result of Bekenstein, we fix the
fundamental scale of the theory. We find that
\begin{equation}
	\label{setscale}
\ell = \frac{1}{2} \sqrt{\frac{\sqrt{3}}{\ln 2}} l_{P}.
\end{equation}
Matching with the thermodynamic result also fixes our fundamental
energy scale.\footnote{Alternately, this can be used to fix the
Immirzi parameter $\gamma$.  One can show that $\gamma = \frac{\ln
2}{\pi \sqrt{3}}$.  Comparing this to the isolated horizons work of
Ref.  \cite{abk}, if the length scale ``$\ell_{P}$'' in Ref.
\cite{abk} is equal to $\sqrt{\hbar G}$, then the result of Eq.
(\ref{setscale}) agrees with that calculation.}

It is also interesting to investigate the mass-temperature relation
for spherical black holes to see whether these solutions also are
found in this class of gravitational statistical mechanics models.
The expected relation is given by
\[
\vev{M} \propto \frac{1}{T}.
\]
Since the energy is simply proportional to the intersection number and
the temperature, it seems that this model appears not to include black
hole solutions of this type.

While the results for spherically symmetric black holes do not match
the thermodynamics of this ensemble, the entropy bound \cite{bentropyb}
\begin{equation}
	\label{entropybound}
	\frac{S}{k} \leq 2 \pi R E
\end{equation}
is satisfied with the areal radius definition $R := \frac{1}{2}
\sqrt{\frac{\vev{A}}{\pi}}$.  To leading order in $\vev{N}$, the
entropy $S$, energy $\vev{E}$ and area $\vev{A}$ scale with $\vev{N}$.
Therefore we have from Eq.  (\ref{entropybound}) that $\vev{N} \leq
\vev{N}^{3/2}$, which is satisfied by all $\vev{N}$; this is not a
tight bound!  The result indicates that the ensembles we have
considered do not contain black holes, which are conjectured to
saturate this bound.

Further refinements of this model, including the implementation of
admissibility conditions due to the topology of the boundary and
gauge invariance, may be found in Ref. \cite{kevinthesis}.

\section{Discussion}
\label{end}

In this work we have shown that a straightforward application of the
ideas of statistical mechanics to a gravitational model yield results
strikingly similar to black hole thermodynamics.  In particular, using
the grand canonical partition function we have shown that in the low
temperature, macroscopic limit the system obeys the Bekenstein-Hawking
entropy relation and the Bekenstein entropy bound.  A new result is
that the system condenses in the lowest level state in the low
temperature, macroscopic limit.

Unlike the previous work in spin net quantum gravity, this work begins
with the quantum gravitational statistical mechanics based on the
Boltzmann weight $e^{-\beta \hat{H}}$.  The pathbreaking work in Refs.
\cite{Rbh,Kbh,abck,abk} based the statistical weight on the concept of
the ``area ensemble,'' in which the Boltzmann weight is based on the
extensive parameter area and its conjugate surface pressure.  Here, we
simply explore the statistical mechanics of the Lorentzian
gravitational model defined in Section \ref{modeldef}.  It is a model
in which the degrees of freedom of quantum geometry are encoded in
non-interacting particles with spin.  We do not specify black hole
or isolated horizon boundary conditions and yet observe much the same
thermodynamics.

There are several technical remarks which ought to be mentioned.  We
remind the reader that the quasilocal operator used in this paper is a
{\em candidate} operator.  In fact, in Ref.  \cite{qqe} there are two
inequivalent expressions for the boundary Hamiltonian corresponding to
two separate normalizations.  In one, the operator is expressed in
terms of the area operator (which provides the quantization of the
$\sqrt{|q_{ab}|}$ in the lapse of Eq.  (\ref{ham})).  This is the
operator used in this paper.  However, the spectrum of the quasilocal
energy operator is modified when vertices lie on the bounding
surface.\footnote{There is an important caveat to this modified
spectrum.  In Ref.  \cite{qqe} the quasilocal energy operates on
states based on networks in the volume and the boundary.  After a
careful study of the quantum boundary conditions and Hilbert space,
the Hamiltonian may operate only on states based on graphs in the
volume.  The area operator in Ref.  \cite{abk} has this property.  If
this is case, then the work done in this paper is far more general.
It would be valid for all vertices, regardless of valence.} The
spectrum for higher valence vertices (type (ii)) is given by
\[
\hat{E}_{\partial \Sigma}({\cal N}) \ket{s}
	=  m \sum_{v \in \partial \Sigma \cap {\mathsf G}}
	{\cal N}_{v}
	\frac{p_{v}(p_{v}+2) + n_{v}(n_{v}+2) - z_{v}(z_{v}+2)}
	{\left[ 2p_{v}(p_{v}+2) + 2n_{v}(n_{v}+2) - z_{v}(z_{v}+2)
	\right]^{\frac{3}{4}}} \ket{s}.
\]
This could be incorporated into the present context by adding another
chemical potential.  Simple statistical arguments suggest that in a
fluctuating surface the dominant contribution would come from the
bivalent intersections used here.

The second operator in Ref.  \cite{qqe} uses a distinct quantization
of the classical Hamiltonian.  The essential difference is that the
determinant of the metric is incorporated using the volume operator.
Both operators are well-defined but on different spaces.  This
operator requires that the intersections between the spin network
state and the boundary are all valence one (type (i)),
meaning that graphs on which states are based cannot have edges
tangent to the surface $\partial \Sigma$.  This spectrum of this
operator is
\begin{equation}
\hat{E}_{\partial \Sigma}({\cal \Ntil} \,) \ket{s_{i}}= m \sum_{v}
{\cal N}_{v} \frac{n_{v}(n_{v}+1)}{\lambda_{v}} \ket{s_{i}},
\end{equation}
where $\lambda_{v}$ is the eigenvalue of the volume
\cite{RSareavol,RC}.  On account of the state restriction, the
numerator always has the simple form of the area operator; the
operator is proportional to $\hat{A_{v}}^{2}/ \hat{V_{v}}$.  The
eigenvalues $\lambda_{v}$ can be computed using recoupling theory
\cite{RC}.  However, since one can always find a surface tangent to a
given edge of an embedded graph, the requirement of ``no tangent
edges'' seems too strict to apply to general boundaries.

Throughout the statistical mechanic computation we set the lapse to 1
on the bounding surface $\partial \Sigma$.  Since the lapse is fixed
and constant by the boundary conditions, our choice is one solution.
This choice, however, reflects the reparameterization freedom of time
and the selection of which temperature we choose to use to investigate
the model.  Thus, the present statistical mechanics is for an observer
at the spatial boundary with unit lapse.  It would be interesting to
investigate a model with another boundary and a parameter which
measured its size so that one could compare solutions as in Ref.
\cite{york}.

We close with three comments on the wider implications of this work.
In the low temperature, macroscopic limit the boundary particles
condense into the lowest level.  Each spin network edge contributes
roughly a Planck area ($\sqrt{3} \ell^{2}$) and has a degeneracy of 2.
In this limit, the model looks like a collection of Planck area
plaquettes which can each be in one of two states, just like a bit.
This is precisely Wheeler's ``It from Bit'' picture of black hole
entropy motivated by information theory
\cite{wheeler}.

In this same limit, variations in fundamental geometry arise from
fluctuations in the number of spin-1/2 edges.  This effectively
reproduces the same parameters as the Bekenstein-Mukhanov model
\cite{BeMu} -- the surface area is a multiple of a fundamental area.
This has far-reaching consequences for the spectrum emitted from the
boundary.  Such a configuration would emit a discrete line spectrum
\cite{Leemog}.  Thus, if edge creation and anhilation is statistically
favored, as our analysis suggests, then the dominant contribution to
radiation would be the Bekenstein-Mukhanov line spectrum.

Finally, it is surprising that such a simple system of independent
``bits of geometry'' has the essential elements to reproduce the
thermodynamics of the gravitational field.  Many other models and
theories have also derived the same thermodynamics.  In string theory,
a 1-dimensional gas of D-branes gives the entropy for extremal black
holes.  In Ref.  \cite{abk} a $U(1)$ Chern-Simons theory gives similar
results for spherically symmetric black holes.  Considering the huge
differences in underlying theories, it seems unavoidable that there is
a strong element of universality of these results.  It appears easy to
derive the thermodynamics of the gravitational field from vastly
different theories of fundamental geometry.  In this sense it seems
that matching thermodynamic results is not a stringent test of quantum
gravity.  With an array of theories giving the ``correct''
thermodynamic limit, we may have to look to new principles to select
the correct quantum theory which underlies general relativity.

\begin{ack}
We thank Mark Taylor for discussions and gratefully acknowledge the
generous hospitality of Deep Springs College and summer research funds
from Swarthmore College.  Thanks also to the relativity and high
energy groups at Syracuse University for their helpful questions and
comments.

\end{ack}

\appendix

\end{document}